\documentclass[%
 reprint,
superscriptaddress,
reprint,
 amsmath,amssymb,
 aps,
prb,
floatfix,
]{revtex4-2}

\usepackage{graphicx}
\usepackage{dcolumn}
\usepackage{bm}
\usepackage{hyperref}
\hypersetup{
    colorlinks=true,
    linkcolor=blue,
    filecolor=blue,      
    urlcolor=blue,
    citecolor=blue
    }

\DeclareMathAlphabet{\mathpzc}{OT1}{pzc}{m}{it}

\usepackage{lipsum}
\usepackage[dvipsnames]{xcolor}
\usepackage{soul}
\newcommand{\p}{\partial}

\usepackage{mathtools}
\usepackage{color}
\usepackage{ulem}

\newcommand{\Rep}{\text{Re}}

\begin{document}


\title{Quantum Hall effect in a chiral cavity}

\author{Liu Yang}
\thanks{These two authors contributed equally}
\affiliation{Tsung-Dao Lee Institute, Shanghai Jiao Tong University, Shanghai, 201210, China}
\affiliation{School of Physics and Astronomy, Shanghai Jiao Tong University, Shanghai 200240, China}
\author{Gabriel Cardoso}
\thanks{These two authors contributed equally}
\affiliation{Tsung-Dao Lee Institute, Shanghai Jiao Tong University, Shanghai, 201210, China}
\affiliation{Nordita, Stockholm University, and KTH Royal Institute of Technology,
Hannes Alfv\'ens v\"ag 12, SE-106 91 Stockholm, Sweden}
\author{Thors Hans Hansson}
\email{hansson@fysik.su.se}
\affiliation{Department of Physics, Stockholm University, AlbaNova University Center, 106 91 Stockholm, Sweden}
\author{Qing-Dong Jiang}
\email{qingdong.jiang@sjtu.edu.cn}
\affiliation{Tsung-Dao Lee Institute, Shanghai Jiao Tong University, Shanghai, 201210, China}
\affiliation{School of Physics and Astronomy, Shanghai Jiao Tong University, Shanghai 200240, China}
\affiliation{Shanghai Branch, Hefei National Laboratory, Shanghai 201315, China}


\begin{abstract} 
We investigate the influence of quantum fluctuations in a chiral cavity on the quantum Hall (QH) state, extending previous studies of QH liquids in linearly polarized cavities. Using the Schrieffer-Wolff transformation for perturbative cavity-matter interaction, we identify the system’s normal modes, which correspond to the elementary excitations of the dressed electrons and photons. In contrast to the linear case, we find that the chiral cavity modifies the Kohn mode frequency by a contribution proportional to the cyclotron frequency, which can be interpreted as a renormalization of the magnetic field by cavity fluctuations. We show that the AC conductivities display cavity-induced corrections, including an isotropic quantum reactance effect and a rotating total-current response under applied AC fields. These findings are also derived from a hydrodynamic approach, which extends their validity to fractional quantum Hall states. Finally, we examine the role of finite cavity quality factor and find that while photon losses introduce resistive contributions to the impedance, these vanish in the DC limit. Our results provide insights into the interplay between quantum Hall states and chiral cavities, with significant implications for material engineering and cavity-induced topological effects.
\end{abstract}

\maketitle


\section{Introduction}

The interplay between vacuum quantum fluctuations and physical systems gives rise to profound phenomena, such as the Casimir effect~\cite{Casimir1948,Casimir1986,Casimir2001,Casimir2004} and the Lamb shift~\cite{Lamb1947,Bethe1947,Welton1948,LambReview2020}. In cavities, these fluctuations can be enhanced by reducing the cavity volume~\cite{cohen,Aspect2010}, making it possible to manipulate materials by vacuum fluctuations in cavities and to probe the strong light-matter coupling regimes~\cite{ultrastrong2012,polariton2019,ultrastrong2019,ultrastrongRMP2019,RevQEDgas2021,Manipulating2021,chiral_cavity2021,review_cavity2022,yaowang2023,lin2023remote}.

Recent studies of quantum Hall (QH) liquids in a cavity have  focused on the robustness of topological material properties in the presence of fluctuating electromagnetic cavity fields~\cite{cavityQHE2022,Ciuti2021,Rokaj2023,Rokaj2024,Ciuti2024,linear2024}. In our associated work~\cite{linear2024}, we established the topological robustness of the QH state from the point of view of QH hydrodynamics. At the same time, we found that the cavity coupling leads to clear dynamical signatures, namely, a modification of the Kohn mode frequency and a quantum reactance effect on the longitudinal conductivity. In the case of integer filling factor, these effects can also be derived from a microscopic model. In this paper, we refine the microscopic approach, and explore additional effects due to the presence of symmetry-breaking in the quantum fluctuations of the cavity field.

Incorporating symmetry breaking into vacuum quantum fluctuations emerges as a powerful method for generating repulsive Casimir forces~\cite{jiang2019chiral,jiang2019axial}, altering spectra~\cite{PhysRevB.99.201104,jiang2023angular}, and engineering material properties within a cavity~\cite{chiralQED2022,lin2025spontaneous}. In particular, chiral cavities can be used to explore the effects of time reversal symmetry (TRS) breaking~\cite{chiralmirrors2015,chiralnano2019,Baranov2020,chiral_cavity2021,chiralQED2022,NCchiralcavity2024,cavitiesTRSB2024}. These include chirality selection in chemical reactions~\cite{jiang2023,chirality_select2024}, flat-band engineering~\cite{jiang2024engineering}, topological phase transitions~\cite{Sentef2019,Ashida2023,jiajun2023,yang2024emergent}, and chiral spin liquid formation~\cite{wei2024cavity}. Since the non-trivial topology of the QH liquids needs TRS breaking~\cite{TKNN1982}, one might wonder if their topological properties are more sensitive to quantum fluctuations in a chiral cavity, and whether there are effects that differ from those observed in the linear case~\cite{cavityQHE2022,Rokaj2023,Rokaj2024,linear2024}. These are the questions that we address in this paper.

Firstly, we use a microscopic description of the integer effect, and discuss the hybridization between the electrons in the lowest Landau level and the cavity mode. By neglecting the spatial variation of the cavity amplitude over the Hall bar, the Galilean invariance is kept, and the center of mass Hamiltonian can be exactly diagonalized~\cite{Kohntheorem,Rokaj2023,Rokaj2024,linear2024}. This is not possible when the amplitude varies over the sample.  Also, it is cumbersome to construct a general diagonalization and calculate the corresponding transport for an arbitrary polarization. Thus, we use to another method, the Schrieffer-Wolff (SW) transformation, which can be applied when the Galilean invariance is broken~\cite{nakajima1955,SWtrans,landi2024,grapheneSW2024}. By a SW transformation, we perturbatively decouple the fermionic and bosonic degrees of freedom and identify two types of elementary excitations: a polariton mode of dressed photons, and a Kohn mode of dressed electrons close to the bare cyclotron frequency. Furthermore, we calculate both the longitudinal and transverse AC conductivities at the center of the QH plateaus \cite{prangebook}. Finally, we analyze the resulting modifications to the longitudinal and Hall resistivities and comment on their relevance for future experiments.

As a complement to the microscopic model, we also examine effects of the chiral cavity on the QH hydrodynamics using the Wen-Zee action, supplemented by higher-derivative (non-topological) terms \cite{zhang1992chern,wen1992shift,[{For the development of the hydrodynamic theory of the QHE, see also  }]stone1990superfluid,*lee1991collective}. The results for both the Hall and longitudinal conductivities, as well as the renormalization of the Kohn mode, are consistent with the microscopic model. While this effective theory does not provide insights into microscopic details, its main advantage lies in its ability to directly generalize to fractional quantum Hall states, for which no analytic microscopic theory currently exists. Our comparison with the microscopic approach also show that, in this formalism, the higher-derivative corrections to the Wen-Zee theory incorporate the contributions of Landau-level transitions to the conductivity. Finally, we consider the effects of photon loss, corresponding to a cavity with finite quality factor~\cite{Aspect2010}, which introduces resistive contributions to the AC conductivities.

The paper is organized as follows. In sections \ref{sec:microscopic}-\ref{sec:sigmaH}, we derive the microscopic model and calculate the observables discussed above. In section \ref{sec:hydro} we give the details of the extended hydrodynamic model, the comparison with the results from the microscopic model in the case of the integer effect, and consider the coupling to a chiral cavity with loss. Finally, in section \ref{sec:conclusions}, we summarize our results and point at possible extensions of our work.

\section{Microscopic model}
\label{sec:microscopic}

\begin{figure}[]
\centering
\includegraphics[width=0.95\linewidth]{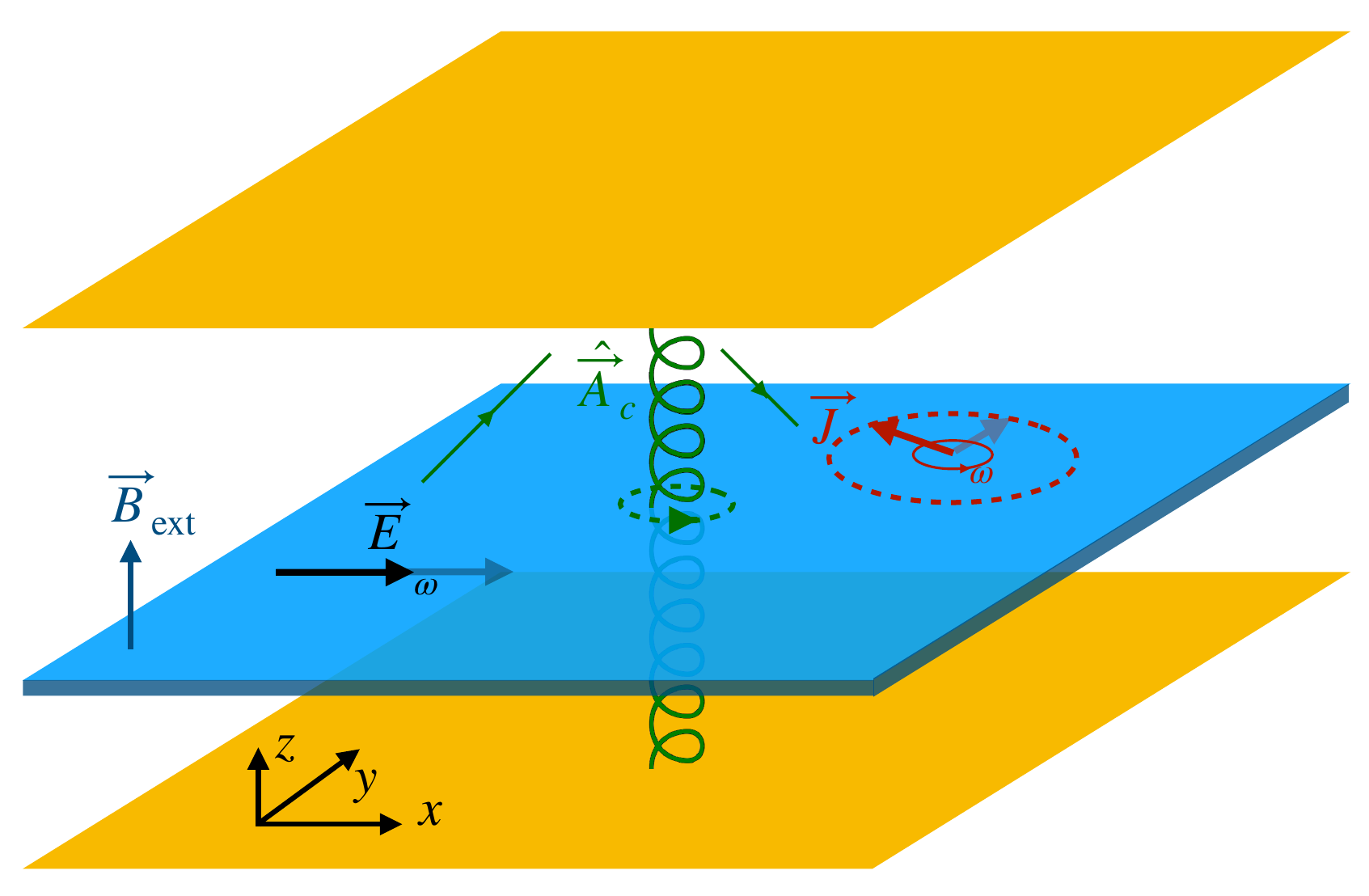}
\caption{The coupling of a Hall bar to a single chiral mode in a cavity can be realized by placing the material between, and parallel to, two special mirrors made by symmetry breaking materials~\cite{chiral_cavity2021}. Second order effects appear in the transport and collective modes, mediated by the exchange of chiral photons with the cavity. In particular, an AC applied field can generate a total current which rotates with the same frequency $\omega$ as the applied field, and the direction of rotation is determined by the chirality of the cavity mode.}
\label{fig:setup}
\end{figure}

We first consider the formation of dressed Landau levels by coupling to a single circularly polarized (chiral) cavity mode~\cite{2018disroder,Rokaj2023}. The Hamiltonian for the $N$ electrons interacting with the cavity field is 
\begin{align}
   H&=\sum_{j=1}^{N}\frac{[\vec{p}_j+e\vec{A}_\text{ext}(\vec{r}_j)+e\hat{\vec{A}}_c]^2}{2m_e}+\omega_\text{c}\left( a^\dagger a+\frac{1}{2}\right),\label{eq:fullH}
\end{align}
where $\vec{p}_j$ are the canonical momenta of the electrons moving in the $x$-$y$ plane, $m_e$ is their effective mass, $\vec{A}_\text{ext}$ is the vector potential corresponding to the external magnetic field $B\,\hat{e}_z=\nabla\times \vec{A}_\text{ext}$, $\omega_\text{c}$ is the frequency of the cavity mode and $\hat{\vec{A}}_c$ is the related field operator. In the chiral cavity, it is quantized as
\begin{align}
\hat{\vec{A}}_c&=A_0\left( \hat{e}_s a+  \hat{e}_s^{\,*} a^{\dagger}\right)\label{eq:A},
\end{align}
with amplitude $A_0=(2\varepsilon\omega_c V_c)^{-1/2}$ ($\varepsilon$ is the relative permittivity of the material), and a complex polarization vector
\begin{align}
  \hat{e}_s=\frac{\hat{e}_y-i s\hat{e}_x}{\sqrt{2}},\label{eq:epsilon_s}
\end{align}
where $s=\pm1$ corresponds to left-/right-handed circular polarization. The photon creation/annihilation operators satisfy the canonical commutation relations $[a,a^\dagger]=1$. Note that we set $c=\hbar=\varepsilon_0=1$, and that we have taken the cavity mode to be spatially uniform. One can visualize this setup in terms of two special mirrors parallel to the Hall bar, which favor a single circularly polarized mode in the enclosed region (see Fig.~\ref{fig:setup}). Such a chiral cavity can be realized in experiments by stacking one Faraday rotator on top of a good mirror~\cite{chiral_cavity2021}, by using flat-band materials~\cite{suarez2024chiral}, or utilizing a magnetoplasma in a doped semiconductor~\cite{cavitiesTRSB2024}. Due to Faraday reflection, the chiral modes with opposite chiralities can have alternating nodal structures. By positioning the two-dimensional material at a plane where one of the chiral modes vanishes, the system can be effectively modeled as a material coupled to a single chiral mode.

The bare electron part of the Hamiltonian describes the Landau levels of the free electron gas, with the cyclotron frequency 
\begin{align}
    \omega_B=\frac{eB}{m_e}.\label{eq:omegab}
\end{align}
In this section, we consider the integer Hall effect, where the electrons fully occupy an integer number $\nu=2\pi N /(e B A_H)$ of Landau levels, with $A_H$ the area of the Hall bar.

The kinetic energy of the electrons includes a term quadratic in the cavity vector potential, which leads to a back reaction on the cavity modes~\cite{back2014,Rokaj2019,CavityQED2021}. For the chiral cavity, this term can be absorbed by a redefinition of the frequency
\begin{align}
    \widetilde{\omega}_c&=\omega_c+ \frac{\alpha \nu}{L_\text{eff}}\frac{ \omega_B}{ \omega_c},\label{eq:tilde_omega}
\end{align}
where $\alpha=e^2/(4\pi)$ is the fine-structure constant and we defined the length scale
\begin{align}
        L_\text{eff}&=\frac{\varepsilon V_c}{A_H},\label{eq:L_eff}
\end{align}
which is determined by the cavity quantization volume $V_c$, the area of Hall bar $A_H$, and the relative permittivity of the material $\varepsilon$. We note that while for an arbitrary polarization the back reaction on the cavity frequency \eqref{eq:tilde_omega} involves also a redefinition of the photonic operators $a,a^\dagger$ and modifies the normal frequency (see Appendix~\ref{app:backreaction}), the photonic part of the Hamiltonian is already diagonal because $\hat{e}_s^{\,2}=0$. With this, the full second-quantized light-matter Hamiltonian becomes
\begin{align}
 H&= H_0+ H',\\
 H_0&=\sum_{n,k_x}\omega_B\left(n+\frac{1}{2}\right) c_{n k_x}^\dagger  c_{n k_x}+\widetilde{\omega}_c\left( a^\dagger a+\frac{1}{2}\right),\\
    H'&=\frac{e\sqrt{N}}{m_e}\vec{\pi}\cdot \hat{\vec{A}}_c,
\end{align}
where we defined the normalized total kinetic momentum operator (for more details, see Appendix \ref{app:SW}),
\begin{align}
   \vec{\pi} &=\sqrt{\frac{eB}{N}}\sum_{nk_x}\sqrt{n+1} c_{n+1k_x}^\dagger  c_{nk_x}\hat{e}_-+\text{h.c.}\, ,\label{eq:pi_vec}
\end{align}
and used the Landau gauge $\vec{A}_\text{ext}=-By\hat{e}_x$ for the background external field, so that the electronic states created by $c_{nk_x}^\dagger$ have a well-defined momentum $k_x$ along the $x$ direction~\cite{prangebook}, with canonical anticommutation relations $\{c_{n k_x},c_{n'k'_x}^\dagger\}=\delta_{nn'}\delta_{k_xk'_x}$. With these definitions, the kinetic momenta \eqref{eq:pi_vec} satisfy the commutation relation 
\begin{align}
      [\pi^y,\pi^x]&=i eB,\label{eq:pi_commu}
\end{align}
when the system has a conserved electron number $N$.

Since the quantum cavity field is approximated as homogeneous, which preserves the Galilean invariance for the whole system, only the center-of-mass of the electrons couples to the cavity, in accordance with Kohn's theorem~\cite{Kohntheorem}. Thus, the Hamiltonian can be separated as a sum of the center-of-mass component,
\begin{align}
    H_\text{CM}&=\frac{(\vec{\pi}+e\sqrt{N}\hat{\vec{A}}_c)^2}{2 m_e}+\omega_c(a^\dagger a+\frac{1}{2}),\label{eq:HCM}
\end{align}
plus a component depending only on the relative distances~\cite{Rokaj2023}. To describe the motion of the center of mass, we introduce the collective coordinate $\vec{R} = \sum_j \vec{r}_j / \sqrt{N}$, in terms of which the momentum operator $\vec{\pi}$ can be written as 
\begin{align}
    \vec{\pi}=-i\nabla_{\vec{R}}+e \vec{A}_{\text{ext}}(\vec{R}).\label{eq:pi_R}
\end{align}
This momentum operator satisfies the commutation relation $[R^i,\pi^j]=i\delta^{ij}$. We will often consider the projection onto the plane wave state $\exp(ik_xR_x)$ for the Landau-gauge external field, which leads to $\vec{\pi}_{k_x}=(k_x-e B R^y)\hat{e}_x-i\partial_{R^y}\hat{e}_y$ and the corresponding Hamiltonian $H_\text{CM}(k_x)$.

\section{Schrieffer-Wolff transformation}
\label{sec:SW}

For weak interaction $H'$, one can describe the system through the perturbative effects of the light-matter coupling on the electronic and photonic states. Our approach is based on the perturbative SW transformation, where a unitary transformation that brings the Hamiltonian to diagonal form is constructed order by order~\cite{nakajima1955,SWtrans,Loss_review,landi2024}.

To first order, one defines the anti-hermitian operator $S$ by the linear equation
\begin{align}
   H'+[ H_0, S]=0,\label{eq:SWeq}
\end{align}
which implies that $S$ is of the order of $H'$. The transformed Hamiltonian $H^S=e^{-S}He^{S}$ is then given by
\begin{align}
   H^S&= H_0+\frac{1}{2}[ H', S]+\mathcal{O}(H'^3).
\end{align}
Note that now the leading correction to the non-interacting Hamiltonian $H_0$ appears only at quadratic order, since $S\sim H'$ by \eqref{eq:SWeq}.

For the chiral cavity, we find the solution
\begin{align}
    S=&\sum_{n,k_x} \sqrt{\frac{\alpha \nu(n+1)}{N L_\text{eff} \omega_c }}c_{n+1k_x}^\dagger c_{nk_x}\bigg(\frac{\delta_{s,+1}a}{\gamma-1}-\frac{i\delta_{s,-1}a^\dagger}{1+\gamma} \bigg)-\text{h.c.},\label{eq:S_chiral}
\end{align}
where the parameter
\begin{align}
    \gamma=\frac{\widetilde{\omega}_c}{\omega_B}=\frac{\omega_c}{\omega_B}+ \frac{\alpha \nu }{L_\text{eff}\omega_c}\label{eq:gammadef}
\end{align}
is assumed to be much smaller than $1$. In Appendix \ref{app:SW}, we derive $S$ in detail for an arbitrary polarization. Applying the $S$ operator \eqref{eq:S_chiral}, we get the effective Hamiltonian for the case of a chiral cavity with polarization $\hat{e}_s$, 
\begin{align}
    H^S&\approx H_0+\frac{\alpha \nu}{L_\text{eff}}\frac{1}{1-s\gamma}\frac{\omega_B}{\omega_c}\bigg(\frac{s\,\vec{\pi}^2}{2eB}- a^\dagger a-\frac{1}{2}\bigg)\, ,\label{eq:HS}
\end{align}
so that, to this order, the SW transformation decouples the system. The Hamiltonian can be split as $H^S=H^S_1+H^S_2$,
\begin{align}
 H_1^S&=\omega_c\bigg(1-\frac{\alpha \nu}{L_\text{eff}}\frac{\gamma}{s-\gamma}\frac{\omega_B}{\omega_c^2}\bigg)( a^\dagger a+\frac{1}{2}),\label{eq:H1}\\
    H_2^S&=\sum_{n,k_x}\omega_B\left(n+\frac{1}{2}\right) c_{nk_x}^\dagger  c_{nk_x}+\frac{\alpha \nu}{L_\text{eff}}\frac{1}{s-\gamma}\frac{\omega_B}{\omega_c}\frac{\,\vec{\pi}^2}{2eB},\label{eq:H2}
\end{align}
where the bosonic Hamiltonian $H_1^S$ and the fermionic Hamiltonian $H_2^S$ describe the dressed cavity photons and dressed electrons respectively.

\section{Dressed excitations}\label{sec:excitations}

The dressed photons, or polaritons, are described by the bosonic Hamiltonian $ H_1^S$. From \eqref{eq:H1}, we directly read out the frequency of the polariton mode,
\begin{align}
    \omega_1&=\omega_c\left(1-\frac{\alpha \nu}{L_\text{eff}}\frac{\gamma}{s-\gamma}\frac{\omega_B}{\omega_c^2}\right)\label{eq:frequency1full}\\
    &\approx \omega_c-s\nu\frac{\alpha}{L_\text{eff}},\label{eq:frequency1}
\end{align}
where $\gamma$ is given by equation \eqref{eq:gammadef}. In the second line, we expanded the correction to the frequency to leading order in the ratios of $\omega,\omega_c$ and $\alpha/L_{\rm eff}$ to $\omega_B$, and in $\alpha/(L_{\rm eff}\omega_c)$. In the remaining of the paper, we will always consider this limit, which corresponds to our regime of interest: widely spaced Landau levels, weakly coupled to the cavity field. As we will discuss later, this is the relevant regime for current experiments. Equation \eqref{eq:frequency1} shows that coupling to the QH system leads to a finite shift of the cavity frequency even in the topological limit $\omega_B\to\infty$, which does not happen in linear cavities~\cite{linear2024}.

In Eq.~(\ref{eq:H2}), the light-matter interaction introduces a term proportional to the square of the total kinetic momentum of the dressed electrons. This term includes an all-to-all mediated interaction between the particles. In the center-of-mass component, it can be combined with the kinetic energy term of the original Landau levels, leading to a modification of the effective mass. Consequently, it results in a modified Kohn mode frequency
\begin{align}
    \omega_2&=\omega_B\left(1+\frac{1}{s-\gamma}\frac{\alpha\nu}{L_\text{eff}\omega_c}\right)\label{eq:frequency2full}\\
    &\approx\omega_B\left(1+s\nu\frac{\alpha}{L_\text{eff}\omega_c}\right)+\nu\frac{\alpha}{L_\text{eff}}.\label{eq:frequency2}
\end{align}
The shift $\nu\alpha/L_\text{eff}$ is identical to the shift observed in the linear cavity case \cite{linear2024}. On the other hand, in contrast to the linear case, we find that in the chiral cavity there is also a multiplicative renormalization of the Kohn mode frequency or, in other words, a shift proportional to $\omega_B$. Since typically $\omega_B\gg\omega_c$, this means that the chiral cavity leads to a much stronger modification of the Kohn mode frequency, which is either enhanced or suppressed depending on the chirality $s$ of the circularly polarized vacuum field. While the $\nu\alpha/L_{\rm eff}$ shift can be understood as a mass renormalization arising a drag effect due to the cavity, it is natural to interpret the $\omega_B$-dependent shift as a renormalization of the magnetic field. Comparing Eq.~(\ref{eq:frequency2}) with \eqref{eq:omegab}, we see that the effective magnetic field generated through the cavity coupling is given by
\begin{equation}
    \delta\vec B=s\nu \frac{\alpha B}{L_\text{eff}\omega_c}\hat{e}_z.\label{eq:deltab}
\end{equation}
We can make this interpretation more rigorous by using a different transformation to eliminate the photon operators from the kinetic energy $(\vec{\pi}+e\sqrt{N}\hat{\vec{A}}_c)^2/(2m_e)$ in the center-of-mass Hamiltonian Eq.(\ref{eq:pi_R}). This is the Power-Zienau-Woolley (PZW) transformation   $U_\text{PZW}=\exp(-ie\sqrt{N}\vec{R}\cdot\hat{\vec{A}}_c)$~\cite{PZW2020,StokesRMP,Stokes2023}. Using $[\hat{A}_c^i,\hat{A}_c^j] = -is A_0^2\epsilon^{ij}$ (note that $\hat{A}_c^i\sim \epsilon^{ij} \hat{E}_{c\,j}$, and $N=\nu eB A_H/2\pi$), we get
\begin{align}
   U_\text{PZW}^\dagger(\vec{\pi}+e\sqrt{N}\hat{\vec{A}}_c )U_\text{PZW}=\vec{\pi}+e\vec{A}_\text{PZW},
\end{align}
with the cavity-induced classical vector potential
 \begin{align}
    \vec{A}_\text{PZW}(\vec{R})=s\nu\frac{\alpha}{2L_\text{eff}\omega_c}B\hat{e}_z\times\vec{R}.\label{eq:Apzw}
\end{align}
This additional potential leads to an emergent magnetic field of the form \eqref{eq:deltab}. Applying the same procedure one finds that, for a linear cavity, the cavity-induced vector potential $\vec{A}_\text{PZW}$ is zero and hence does not give rise to any renormalization of the magnetic field.

The frequency shifts of the dressed normal modes (\ref{eq:frequency1full}) and (\ref{eq:frequency2full}) are not particular to the quantum Hall system, but rely only on the Galilean invariance of the two-dimensional electron gas in spatially constant background magnetic and cavity fields. This is illustrated in  Appendix.~\ref{app:exact_check}, where we go beyond the perturbative SW approach and directly diagonalize the Hamiltonian for a constant left-handed cavity mode coupled to the CM degrees of freedom for the electron gas.

\section{Transport properties}
\label{sec:sigmaH}
We use the Kubo formalism to calculate the AC conductivities for a Hall bar in a chiral cavity~\cite{kubo1965,Rokaj2023}. Since the cavity mode is circularly polarized, the effect of cavity fluctuations is isotropic, so that the conductivity tensor has the form
\begin{equation}
    \sigma^{ij}=\sigma_L\delta^{ij}+\sigma_H\epsilon^{ij}.\label{eq:sigma_tensor}
\end{equation}
The isotropy stems from the z-axis rotational symmetry of the system (see Fig.~\ref{fig:setup}). The longitudinal AC conductivity and is calculated as
\begin{align}
    \sigma_L(\omega)&=\frac{e^2\nu}{2\pi}\sum_{\mathbf{m}} i \omega\frac{2E_{\mathbf{m}\mathbf{0}}}{E_{\mathbf{m}\mathbf{0}}^2-\omega^2}|\langle\mathbf{0}|\frac{\pi^{x}_{k_x}}{\sqrt{eB }}|\mathbf{m}\rangle|^2,\label{eq:Kubo_aa}
\end{align}
where $|\mathbf{m}\rangle$ and $E_{\mathbf{m}}$ are the eigenstates and eigenvalues of $H_{\text{CM}}(k_x)$, and $E_{\mathbf{m}\mathbf{0}}=E_{\mathbf{m}}-E_{\mathbf{0}}$. For chiral cavities, the AC Hall conductivity simplifies to
\begin{align}
     \sigma_H(\omega)&=\frac{e^2\nu}{2\pi}\bigg[1+\frac{2\omega^2}{eB}\sum_{\mathbf{m}}\frac{\text{Im}(\langle\mathbf{0}|\pi^{y}_{k_x}|\mathbf{m}\rangle\langle\mathbf{m}|\pi^{x}_{k_x}|\mathbf{0}\rangle)}{E_{\mathbf{m}\mathbf{0}}^2-\omega^2}\bigg].\label{eq:Kubo_xy}
\end{align}
Due to the homogeneity of the system, the Hall conductivity is equal to the Hall conductance. Thus, for the Hall transport, we shall not distinguish them.

\subsection{Calculation in SW-transformed representation}
Since the light-matter interaction is approximately decoupled in Eq.~(\ref{eq:HS}), the eigenstates of the center-of-mass Hamiltonian in the SW-transformed representation can be constructed as
\begin{align}
    |\mathbf{m}=(m_1,m_2)\rangle^S=\frac{ a^{\dagger m_1} b_{k_x}^{\dagger m_2}}{\sqrt{m_1!m_2!}}|0_a\rangle|0_b\rangle\, ,
\end{align}
where we introduced the Landau level ladder operators
\begin{align}
    b_{k_x}=\frac{\pi_{k_x}^y+i\pi_{k_x}^x}{\sqrt{2 eB}},\quad   b_{k_x}^\dagger=\frac{\pi_{k_x}^y-i\pi_{k_x}^x}{\sqrt{2 eB}},
\end{align}
which satisfy the canonical commutation relation $[b_{k_x},b_{k_x}^\dagger]=1$, and the states $|0_a\rangle$ and $|0_b\rangle$ satisfy $ a|0_a\rangle =b_{k_x}|0_b\rangle = 0$. By using the $S$ operator in Eq.~(\ref{eq:S_chiral}), we can express the different components of the SW-transformed operator $\vec{\pi}_{k_x}$ as follows
\begin{align}
    \pi^{Sx}_{k_x}&=\sqrt{\frac{eB}{2}}i(b_{k_x}^\dagger-b_{k_x})-i\sqrt{\frac{\alpha \nu}{L_\text{eff}\omega_c}}\frac{\sqrt{e B } }{s-\gamma}\frac{a^\dagger-a}{\sqrt{2}},\\
    \pi^{Sy}_{k_x}&=\sqrt{\frac{eB}{2}}(b_{k_x}^\dagger+b_{k_x})-s\sqrt{\frac{\alpha \nu}{L_\text{eff}\omega_c}}\frac{\sqrt{e B }}{s-\gamma}\frac{a^\dagger+a}{\sqrt{2}}.
\end{align}

By applying the transformed states and operators, we find that to leading order, the AC conductivities are given as
\begin{align}
    \sigma_L(\omega)&=\nu\sigma_0\bigg(\frac{i\omega \omega_B}{\omega_B^2-\omega^2}+\frac{\alpha \nu}{L_\text{eff}}\frac{i\omega}{\omega_c^2-\omega^2}\bigg),\label{eq:ACL}\\
     \sigma_H(\omega)&=\nu\sigma_0\left(1+\frac{\omega^2}{\omega_B^2-\omega^2}+s\frac{\alpha \nu}{L_\text{eff}\omega_c}\frac{\omega^2}{\omega_c^2-\omega^2}\right),\label{eq:ACH}
\end{align}
where $\sigma_0=e^2/(2\pi)$ is the conductance quantum. Both the longitudinal and Hall conductivities have cavity-induced modifications, with poles at \(\pm \omega_c\), while the poles at \(\pm \omega_B\) correspond to transitions between Landau levels. At low frequencies, these terms contribute linearly to \(\sigma_L(\omega)\), but quadratically to \(\sigma_H(\omega)\). For the left-handed cavity (\(s=1\)), we have verified the results in Eqs.~(\ref{eq:ACL}) and (\ref{eq:ACH}) against the exact calculations in Appendix~\ref{app:exact_check}. Furthermore, the chirality of the cavity mode determines the sign of the term with poles at $\pm\omega_c$ in the AC Hall conductance. When $s=1$ $(-1)$, the cavity enhances (suppresses) the Hall conductance at low frequencies. 

\subsection{Topological limit}\label{sec:qreact}

To discuss the topological properties of a system, we must work in the ``topological limit" (for a definition see section \ref{sec:hydro}) and in our case it amounts to $\omega_B\to\infty$ implying    
\begin{align}
    &\sigma_L=\nu\sigma_0\frac{\nu\alpha}{L_{\rm eff}}\frac{i\omega}{\omega_c^2-\omega^2},\label{eq:sigmaLtop}\\
    &\sigma_H=\nu\sigma_0\left(1+s\frac{\nu\alpha}{L_{\rm eff}\omega_c}\frac{\omega^2}{\omega_c^2-\omega^2}\right).\label{eq:sigmaHtop}
\end{align}
The correction to the longitudinal conductivity is given by a quantum reactance term, which we discussed in \cite{linear2024} for the case of linear polarization. It describes the dissipasionless exchange of energy between the electrons in the QH liquid and the cavity photons, and in the chiral case it appears isotropically in the longitudinal conductivity. Additionally, the exchange of chiral photons with the cavity generates not only a longitudinal reactive current, but also a transverse current. The extra AC Hall contribution is also a dissipationless effect, and remains non-vanishing in the topological limit.

Considering both the longitudinal and the transverse effects, one finds that a rotating current response is generated by an applied AC linearly-polarized electric field
\begin{equation}
    \vec E(t)=E_0\cos(\omega t)\vec e_x.
\end{equation}
Indeed, from Eqs (\ref{eq:sigmaLtop},\ref{eq:sigmaHtop}), one finds that the cavity-induced AC response current is
\begin{equation}
    \delta\vec J=-\frac{\alpha\nu^2\sigma_0E_0}{L_{\rm eff}}\frac{\omega}{\omega_c^2-\omega^2}[\sin(\omega t)\vec e_x+s\frac{\omega}{\omega_c}\cos(\omega t)\vec e_y],
\end{equation}
which describes an elliptical contour as a function of time, with orientation determined by the chirality $s$ of the cavity mode (see Fig.~\ref{fig:setup}). Note that the power density, $\vec E\cdot \delta \vec J$ varies periodically in time and integrates to zero over a period.

\begin{figure}[]
\centering
\includegraphics[width=0.95\linewidth]{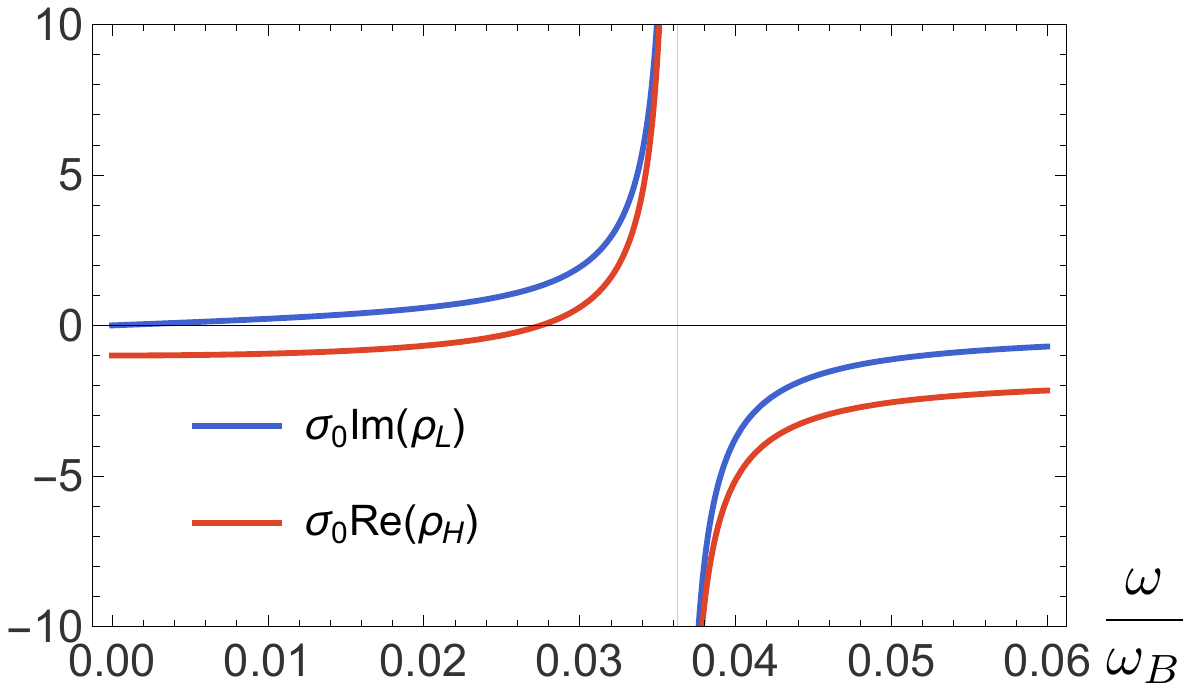}
\caption{Leading order AC Resistivities for QH liquids in a left-handed ($s=+1$) chiral cavity in the topological limit given in Eqs.~(\ref{eq:rho_L},\ref{eq:rho_H}). The unit of the Resistivities is $\sigma_0^{-1}=2\pi/e^2$ and the frequency is in units of the cyclotron frequency $\omega_B$. The parameters used here are the same as in our associated work\cite{linear2024}, with a sample area of $A_H = 40 \times 200\mu\text{m}^2$, filling factor $\nu = 1$, cyclotron frequency $\omega_B = 24.3\text{THz}$, cavity frequency $\omega_c = 0.88\text{THz}$, and effective length $L_\text{eff} = 3.38\mu\text{m}$. The polarization is set to be $\hat{e}_+$. At low frequencies, $\rho_L$ and $\rho_H$ have different power dependencies on the frequency. Around the pole $\omega = \omega_c$, we have $\rho_L \approx is \rho_H$.}
\label{fig:plotrhoQ0}
\end{figure}

Inverting the finite-$\omega_B$ expressions (\ref{eq:ACL},\ref{eq:ACH}) and again expanding to leading order in $1/\omega_B$ and the weak coupling to the cavity field, we obtain, for the resistivity tensor $\rho^{ij}=\rho_L\delta^{ij}+\rho_H\epsilon^{ij}$,
\begin{align}
    &\rho_L=\frac{i\sigma_0^{-1} \alpha}{ L_{\rm eff}\omega_c}\frac{\omega}{\omega_c^2-\omega^2},\label{eq:rho_L}\\
    &\rho_H=(\nu \sigma_0)^{-1}\left(-1+s\nu\frac{\alpha}{L_{\rm eff}\omega_c}\frac{\omega^2}{\omega_c^2-\omega^2}\right).\label{eq:rho_H}
\end{align}
These formulas can be directly compared with experiments. We see that the longitudinal component $\rho_L$ indeed corresponds to a purely reactive impedance, and in the DC limit one recovers the standard values $\rho_L\to 0$, $\rho_H\to (\nu \sigma_0)^{-1}$ at the center of the Hall plateau. In Fig.~\ref{fig:plotrhoQ0}, we show $\rho_L$ and $\rho_H$ as a function of $\omega/\omega_B$ in unit of $\sigma_0$ for a Hall bar in a left-handed ($s=+1$) cavity. At low frequencies, $\rho_L$ is linear in $\omega$, while $\rho_H$ quadratic in $\omega$. Close to the pole $\omega = \omega_c$, the cavity correction dominates the conductivities, and one finds that $\rho_L \approx i s\rho_H\sim(\omega_c^2-\omega^2)^{-1}$.

\section{Hydrodynamic formulation}
\label{sec:hydro}

An alternative formulation, which is useful in bringing out the universal features of the response to cavity fluctuations, is the hydrodynamics of the QH liquid. Consider the Wen-Zee hydrodynamic action,
\begin{align}
    {\cal L}_H= -\frac m {4\pi} \epsilon^{\mu\nu\sigma}b_\mu\partial_\nu b_\sigma-\frac e {2\pi} \epsilon^{\mu\nu\sigma}(A+A_c)_\mu\partial_\nu b_\sigma\nonumber\\
    + \frac m {4\pi \omega_B }\vec E_b^2 - \frac u 2 B_b^2,\label{eq:hydroL}
\end{align}
where we have separated the probe electromagnetic field $A$ from the cavity fluctuations $A_c$. This action describes both the integer $m=1$ and fractional $\nu=1/m$ QH liquids. The hydrodynamic field $b$ parametrizes the electron current as $J^\mu=\frac 1 {2\pi} \epsilon^{\mu\nu\sigma}\partial_\nu b_\sigma$. The quadratic terms in the combinations $\vec E_b^i = -\partial_t \vec b^i - \partial^i b^0$ and $B_b = \epsilon^{ij}\partial_i b_j$ are the first sub-leading corrections to the Wen-Zee action in the derivative expansion of the hydrodynamics. The constant $u$ corresponds to a delta function repulsive potential, but one can also consider the more general case of a non-local potential $u(|\vec r_1 -\vec r_2|)$. The topological limit  amounts to neglecting the higher derivative terms in \eqref{eq:hydroL}, i.e. to $\omega_B \rightarrow \infty$, and $u\rightarrow 0$.

We use the single-mode approximation $\vec{A}_c(t,\vec{x})=\Rep[q(t)\phi_{k_c}(\vec{x})\hat e_s]$, with the circular polarization \eqref{eq:epsilon_s}. The mode function satisfies $\Delta\phi_{k_c}=-k_c^2\phi_{k_c}$, the boundary conditions in the cavity, and the normalization $\int_{V_c} d^3x\phi_{k_c}=1$. This definition generalizes the previous expression \eqref{eq:A} to non-homogeneous cavity modes. The fluctuations of the complex amplitude $q(t)=q_1+iq_2$ are described by the single-mode Lagrangian,
\begin{equation}
    L_c = \frac{\varepsilon\omega_c}{2}q^*(i\partial_t-\omega_c)q,\label{eq:Lccomplex}
\end{equation}
as reviewed in Appendix~\ref{app:hydro}, and the cavity frequency is given by $\omega_c = \frac{k_c}{\sqrt{\varepsilon\mu}}$. Integrating out the cavity mode amplitude fluctuations results in the contribution
\begin{align}
    &\frac{1}{\varepsilon}\left(\frac{e}{4\pi}\right)^2\int\frac{d\omega}{2\pi}\Big[\frac{\omega^2 \,\delta^{ij}}{\omega_c^2-\omega^2}+\frac{is\omega^3 \, \epsilon^{ij}}{\omega_c(\omega^2-\omega_c^2)}\Big]b_{c,i}^\ast b_{c,j}(\omega),\label{eq:chiralterm}
\end{align}
to the hydrodynamic action. Here, we use the notation $f_c(t) = \int d^2x\,\phi_{k_c}|_H f(t,\vec x)$ for the weighted average of a function on the Hall bar, with weights given by the mode function of the cavity, $\phi_{k_c}|_H$, and we chose the radiation gauge, $v^0=\vec\nabla\cdot\vec v=0$, for both the cavity and the hydrodynamic fields. In contrast with the linear cavity \cite{linear2024}, the resulting contribution is isotropic (see also Appendix \ref{app:hydro}). Note that the sign of the chiral contribution, $\propto\epsilon^{ij}$, is given by the chirality of the cavity mode $s$. We now discuss how this term modifies the properties of the QH liquid.

\subsection{Comparison with the microscopic approach}

To compute the shift in the Kohn mode frequency, let us take the case of homogeneous $b$ and zero probe field. In the radiation gauge, the coupling term becomes $\frac{e}{2\pi}A_H\int\frac{d\omega}{2\pi}i\omega\epsilon^{kl}b_k(\omega) A_{c,l}(-\omega)$, and the full hydrodynamic action simplifies to
\begin{align}
    &\int\frac{d\omega}{2\pi}\frac{A_H}{4\pi}\Big\{\left[im\omega\epsilon^{ij}+\frac{m}{\omega_B}\omega^2\delta^{ij}\right]b_i^\ast b_j(\omega)\label{eq:bareKohn}\\
    &+\frac{\alpha}{L_{\rm eff}}\frac{\omega^2}{\omega_c(\omega^2-\omega_c^2)}\left[is\omega\epsilon^{ij}-\omega_c\delta^{ij}\right]b_{i}^\ast b_{j}(\omega)\Big\}\nonumber.
\end{align}

The saddle point equations have the form $M^{ij}(\omega)b_j=0$, which have nonzero solutions at the frequencies determined by the characteristic equation $\det M(\omega)=0$. For a matrix of the form $M^{ij}=c\delta^{ij}+id\epsilon^{ij}$, the solution is given by $c=\pm d$. Thus, in the absence of coupling, the first line \eqref{eq:bareKohn} gives the bare value of the Kohn mode frequency $\omega=\pm\omega_B$. Considering the full matrix, we find the leading-order corrections to the Kohn mode frequency,
\begin{equation}
    \omega = \omega_B\left(1+s\frac{1}{m}\frac{\alpha}{L_{\rm eff}\omega_c}\right)+\frac{1}{m}\frac{\alpha}{L_{\rm eff}}.\label{eq:kohnhydro}
\end{equation}
For the case of integer filling $\nu=1/m=1$, this expression agrees with the result obtained from the microscopic calculation, Eq.~\eqref{eq:frequency2}.

Next, consider the contribution of the cavity modification \eqref{eq:chiralterm} to the transport. For simplicity, we consider the effective action $S_{\rm eff}[A]$ for homogeneous probe field. 
Integrating the $b$ field by using the matrix inversion formula $(c\delta^{ij}+id\epsilon^{ij})^{-1}=(c^2-d^2)^{-1}(c\delta^{ij}-id\epsilon^{ij})$, we find
\begin{align}
    &S_{\rm eff}=-\frac{e^2A_H}{4\pi m}\int\frac{d\omega}{2\pi}\bigg[\bigg(\frac{\omega^2\omega_B}{\omega^2-\omega_B^2}+\frac{1}{m}\frac{\alpha}{L_{\rm eff}}\frac{\omega^2}{\omega^2-\omega_c^2}\bigg)\delta^{ij}\nonumber\\
    &\hspace{.05\linewidth}-i\left(\frac{\omega\omega_B^2}{\omega^2-\omega_B^2}+s\frac{1}{m}\frac{\alpha}{L_{\rm eff}\omega_c}\frac{\omega^3}{\omega^2-\omega_c^2}\right)\epsilon^{ij}\bigg]A_iA_j
\end{align}
up to higher-order terms. Varying over the probe field, we find
\begin{align}
    &\frac{\sigma_L}{e^2/(2\pi m)}=\frac{i\omega\omega_B}{\omega_B^2-\omega^2}+i\frac{1}{m}\frac{\alpha}{L_{\rm eff}}\frac{\omega}{\omega_c^2-\omega^2},\label{eq:hydrosigmaxxfull}\\
    &\frac{\sigma_H}{e^2/(2\pi m)}=1+\frac{\omega^2}{\omega_B^2-\omega^2}+s\frac{1}{m}\frac{\alpha}{L_{\rm eff}\omega_c}\frac{\omega^2}{\omega_c^2-\omega^2},\label{eq:hydrosigmaxyfull}
\end{align}
which in the case of integer filling fraction, $\nu=1/m=1$, agree with the microscopic calculation (\ref{eq:ACL},\ref{eq:ACH}). In particular, we see that the contributions from the Landau-level transitions are captured by the higher-order terms in the hydrodynamic action \eqref{eq:hydroL}, with the topological limit discussed in section \ref{sec:qreact} following from the cavity coupling to the Wen-Zee hydrodynamics. Finally, we note that the hydrodynamic calculation does not only confirm the microscopic results for the Kohn mode and conductivity, but also extends our conclusions to the fractional quantum Hall states.

\subsection{Finite $Q$ cavity}
\label{sec:qcavity}

It is also simple to extend the hydrodynamic calculation to incorporate the effect of loss in the cavity. An effective description is given by the cavity mode Lagrangian
\begin{equation}
    L_c = \frac{\varepsilon\omega_c}{2}q^*(i\partial_t-\omega_c+i\kappa)q,\label{eq:Lagrangiankappa}
\end{equation}
where $\kappa$ is the loss rate of cavity photons. This effective, non-Hermitian description, follows from coupling the cavity mode to a bath of simple harmonic oscillators, as in the Caldeira-Legget model \cite{caldeira1981influence,wilkinson1990dissipation}. The spectral function,
\begin{equation}
    -\frac{1}{\pi}\text{Im}\, G^R(\omega) = \frac{1}{\pi}\frac{\kappa}{(\omega-\omega_0)^2+\kappa^2},
\end{equation}
where $G^R$ is the retarded Green's function corresponding to the Lagrangian \eqref{eq:Lagrangiankappa}, implies a frequency broadening $\Delta\omega=2\kappa$, and we define the quality factor
\begin{equation}
    Q = \frac{\omega_c}{\Delta\omega} = \frac{\omega_c}{2\kappa}.
\end{equation}

Repeating the steps of the previous subsection, we get modified expressions for $\sigma_L=\sigma^{xx}=\sigma^{yy}$ and $\sigma_H=\sigma^{xy}$,
\begin{align}
    &\frac{\sigma_L}{e^2/(2\pi m)}=i\frac{1}{m}\frac{\alpha}{L_{\rm eff}}\frac{\omega}{\omega_c^2-\omega^2}-\frac{1}{m}\frac{\kappa}{\omega_c}\frac{\alpha}{L_{\rm eff}}\frac{\omega(\omega_c^2+\omega^2)}{(\omega_c^2-\omega^2)^2},\nonumber\\
    &\frac{\sigma_H}{e^2/(2\pi m)}=1+s\frac{1}{m}\frac{\alpha}{L_{\rm eff}}\frac{\omega^2}{\omega_c(\omega_c^2-\omega^2)}\\
    &\hspace{.4\linewidth}-2is\frac{1}{m}\frac{\kappa}{\omega_c}\frac{\alpha}{L_{\rm eff}}\frac{\omega^2\omega_c}{(\omega_c^2-\omega^2)^2},\nonumber
\end{align}
where we also expanded in the loss rate $\kappa$ and took the topological limit $\omega_B\to\infty$. Recall that, in the $\kappa\to 0$ limit, the longitudinal response has no dissipation, since the conductivity is imaginary, leading to a $\pi/2$ phase lag between the current and the applied field. This is not true for finite $\kappa$, which introduces a real contribution to the longitudinal conductivity. 

\begin{figure}[]
\centering
\includegraphics[width=0.95\linewidth]{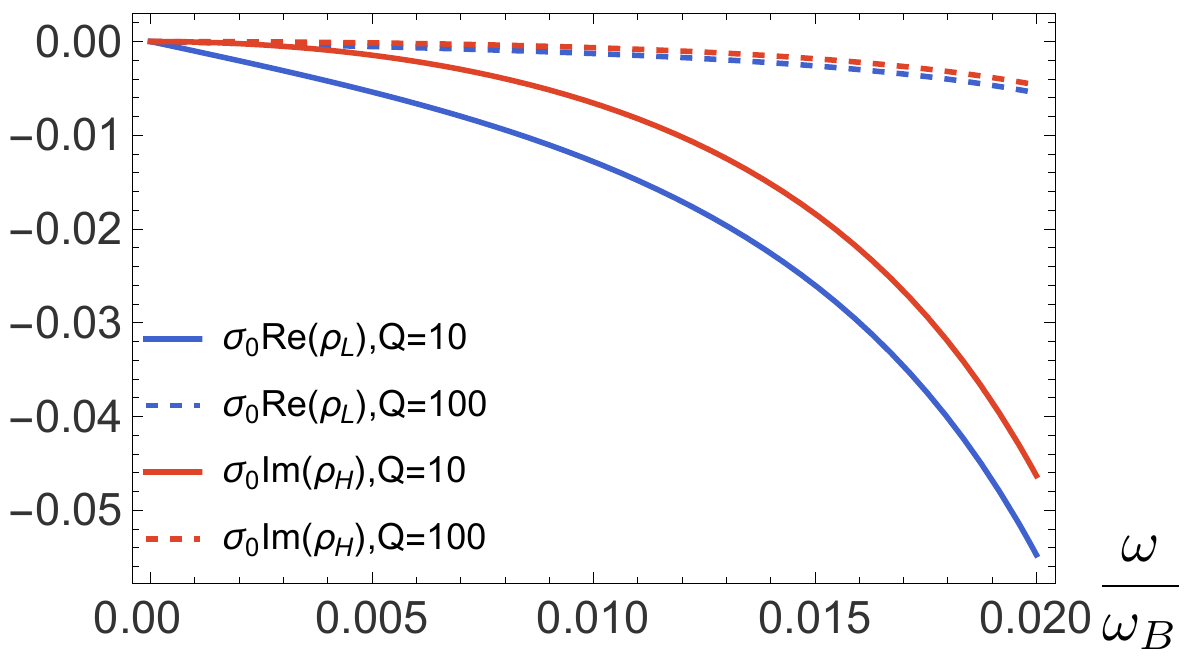}
\caption{Resistivities in units of $\sigma_0^{-1}$ induced by the loss of a left-handed chiral cavity with a quality factor $Q$ (see Eqs.(\ref{eq:rhoL_k},\ref{eq:rhoH_k})). Here, we show the examples with $Q=10,100$. The parameters except for $Q$ are shown in the caption of Fig.~\ref{fig:plotrhoQ0}.}
\label{fig:plotrhoQ}
\end{figure}

By the same derivation as for Eqs. (\ref{eq:rho_L},\ref{eq:rho_H}), we find the resistivities 
\begin{align}
    &\frac{\rho_L}{2\pi m/e^2}=i\frac{1}{m}\frac{\alpha}{L_{\rm eff}}\frac{\omega}{\omega_c^2-\omega^2}-\frac{1}{m}\frac{\alpha\kappa}{L_{\rm eff}\omega_c}\frac{\omega(\omega_c^2+\omega^2)}{(\omega_c^2-\omega^2)^2},\label{eq:rhoL_k}\\
    &\frac{\rho_H}{2\pi m/e^2}=-1+\frac{s\alpha}{m L_{\rm eff}\omega_c}\frac{\omega^2}{\omega_c^2-\omega^2}-\frac{2is\alpha\kappa}{mL_{\rm eff}}\frac{\omega^2}{(\omega_c^2-\omega^2)^2}.\label{eq:rhoH_k}
\end{align}
The cavity loss leads to a resistive contribution to the longitudinal impedance, even at the center of the plateaux. Still, this contribution vanishes in the DC limit, which is a consequence of the single-mode approximation. The energy loss only happens through the exchange of cavity photons, which can leak out. This process is most effective when the applied field has frequency close to resonance with the cavity. Interestingly, the finite $\kappa$ also leads to an out-of-phase component in the transverse impedance, which can in principle also be extracted from AC measurements. In Fig.~\ref{fig:plotrhoQ}, we show the additional resistivities at low frequency region induced by the loss of a chiral cavity with $s=+1$ and $Q=10,100$.

\section{Discussion and Conclusions}
\label{sec:conclusions}

We have explored the impact of chiral cavity fluctuations on the quantum Hall (QH) state, extending previous studies of QH liquids in linear cavities~\cite{Rokaj2023,Rokaj2024,linear2024}. By applying the Schrieffer-Wolff (SW) transformation, we determined the normal modes of the system, which correspond to the elementary excitations of the dressed electrons and photons. These excitations have renormalized frequencies that differ from the bare cyclotron and cavity frequencies. We found that the chiral cavity leads to an extra modification of the Kohn mode by a contribution proportional to the cyclotron frequency $\omega_B$, which is not present in the linear cavity. This contribution can be interpreted as a renormalization of the magnetic field by the cavity fluctuations. For the transport properties, we find that both the longitudinal and transverse AC conductivities have measurable cavity-induced corrections, even in the topological limit, giving experimentally observable effects that differ from those in linear cavities~\cite{linear2024}. As in the linear case, the cavity induces a longitudinal quantum reactance effect, with the difference that in the chiral case this effect is isotropic. Additionally, the cavity also generates a transverse effect, which lead to a rotating total current in response to applied AC fields. For comparison, the features of the QH state in chiral and linear cavities are summarized in Table~\ref{table:comparison} in Appendix~\ref{app:hydro}. We supplemented our discussion by a hydrodynamic calculation, which extends the validity of these results to fractional quantum Hall states. Our calculation also shows that the higher-gradient corrections to the Wen-Zee hydrodynamics correctly reproduce the contributions generated by Landau-level transitions. Finally, we considered the modifications of these results in a finite-$Q$ cavity. We found that, while the cavity loss leads to resistive contributions to the impedance, these contributions vanish in the DC limit. Together, our findings provide a deeper understanding of the interplay between quantum Hall states and chiral cavities, with important implications for future research on topological matter and cavity-induced effects.

\section{Acknowledgments}

THH thanks the director and members of the T.D. Lee Institute for kind hospitality during October of 2024. We gratefully acknowledge previous helpful discussions and feedback from Qian Niu, Kun Yang, and Alexander Abanov. This work was supported by National Natural Science Foundation of China (NSFC) under Grant No. 12374332,  Jiaoda 2030 program WH510363001-1, the Innovation Program for Quantum Science and Technology Grant No. 2021ZD0301900.

\appendix
\section{Back reaction on the cavity mode}\label{app:backreaction}
The back reaction of matter modifies the normal modes of the photons in a cavity~\cite{CavityQED2021,Rokaj2019}. In the following, we will show how to diagonalize the cavity modes by including the back reaction. 

The Hamiltonian for photons in Eq.~(\ref{eq:fullH}) is
\begin{align}
    H_p=\frac{Ne^2\hat{\vec{A}}_c^2}{2m_e}+\omega_c(a^\dagger a+\frac{1}{2}).\label{eq:H_p0}
\end{align}
For a general vector potential for a single uniform cavity mode
\begin{align}
    \hat{\vec{A}}_c=A_0(\hat{e}_c a+\hat{e}_c^{\,\ast} a^\dagger),
\end{align}
we have
\begin{align}
    H_p&=(\omega_c+\frac{\alpha \nu \omega_B}{L_\text{eff}\omega_c})(a^\dagger a+\frac{1}{2})+\frac{\alpha \nu \omega_B}{2L_\text{eff}\omega_c}(\hat{e}_c^{\,2}a^2+\hat{e}_c^{\,\ast2}a^{\dagger 2}),
\end{align}
where we introduced a unit complex vector $\hat{e}_c$ ( $\hat{e}_c\cdot\hat{e}_c^{\,\ast}=1$). Note that $\hat{e}_c$ can be an arbitrary polarization vector, while $\hat{e}_s$ defined in Eq.~(\ref{eq:epsilon_s}) is used for the circular polarizations. Generally, $\hat{e}_c^{\,2}$ is a complex number, 
\begin{align}
   \hat{e}_c^{\,2}=|\hat{e}_c^{\,2}|e^{i\varphi_c}.
\end{align}
Under the transformation $a\to a\exp(-i\varphi_c/2)$, $a^\dagger\to a^\dagger\exp(i\varphi_c/2)$ and $\hat{e}_c\to\hat{e}_c\exp(i\varphi_c/2)$, $\hat{e}_c^{\,2}$ can be made real, while the vector potential remains invariant. Therefore, we can restrict our analysis to cases where $\hat{e}_c^{\,2}$ is real. This condition imposes that the polarization vector $\hat{e}_c$ satisfies
\begin{align}
    \text{Arg}(\hat{e}_c\cdot\hat{e}_+)=-\text{Arg}(\hat{e}_c\cdot\hat{e}_-),
\end{align}
due to the relation $\hat{e}_c^{\,2}=2(\hat{e}_c\cdot\hat{e}_+)(\hat{e}_c\cdot\hat{e}_-)$. To diagonalize $H_p$, we introduce the canonical operators for the cavity field:
\begin{align}
  u=\frac{a^\dagger+a}{\sqrt{2}},\quad v=i\frac{a^\dagger-a}{\sqrt{2}},
\end{align}
which satisfy the commutation relation $[u,v]=i$. In terms of $u$ and $v$, the photonic Hamiltonian is diagonalized as follows
\begin{align}
    H_p&=\frac{1}{2}\bigg[(\widetilde{\omega}_c+ \frac{ \alpha \nu\omega_B \hat{e}_c^{\,2}}{ L_\text{eff}\omega_c})u^2+(\widetilde{\omega}_c-\frac{ \alpha \nu\omega_B \hat{e}_c^{\,2}}{ L_\text{eff}\omega_c})v^2\bigg]\nonumber\\
    &=\omega_p\big(a_p^\dagger a_p +\frac{1}{2}\big),\label{eq:Hp_normal}\\
    a_p&=\bigg(\frac{\widetilde{\omega}_c+ \frac{ \alpha \nu\omega_B \hat{e}_c^{\,2}}{ L_\text{eff}\omega_c}}{\widetilde{\omega}_c- \frac{ \alpha \nu\omega_B \hat{e}_c^{\,2}}{ L_\text{eff}\omega_c}}\bigg)^{1/4}\frac{u}{\sqrt{2}}+\bigg(\frac{\widetilde{\omega}_c- \frac{ \alpha \nu\omega_B \hat{e}_c^{\,2}}{ L_\text{eff}\omega_c}}{\widetilde{\omega}_c+ \frac{ \alpha \nu\omega_B \hat{e}_c^{\,2}}{ L_\text{eff}\omega_c}}\bigg)^{1/4}\frac{i v}{\sqrt{2}},
\end{align}
where  $\widetilde{\omega}_c=\omega_c+ \alpha \nu\omega_B/(L_\text{eff}\omega_c)$ (see $\widetilde{\omega}_c$ in Eq.~(\ref{eq:tilde_omega})), and the normal frequency $ \omega_p$ of $H_p$ can be expressed as
\begin{align}
    \omega_p=\sqrt{\bigg[\omega_c+ \frac{\alpha \nu}{L_\text{eff}}\frac{ \omega_B(1+\hat{e}_c^{\,2})}{ \omega_c}\bigg]\bigg[\omega_c+ \frac{\alpha \nu}{L_\text{eff}}\frac{ \omega_B(1-\hat{e}_c^{\,2})}{ \omega_c}\bigg]}.\label{eq:Omega_c}
\end{align}
It can be proven that $\hat{e}_c^{\,2}$ is always smaller than 1. When $\hat{e}_c=\hat{e}_s$ (see $\hat{e}_s$ in Eq.~(\ref{eq:epsilon_s})), $\omega_p$ is equal to $\widetilde{\omega}_c$. For a linear polarization, such as $\hat{e}_c=\hat{e}_{y}$, we have $\omega_p=\sqrt{\omega_c^2+ 2\alpha \nu \omega_B/L_\text{eff}}$~\cite{linear2024}.

\section{Schrieffer-Wolff transformation}\label{app:SW}

\subsection{$S$ operator}
In the following content, we will show the details of the SW transformation discussed in Sec.~\ref{sec:SW}. 

First, we derive the operator $\vec{\pi}$ in Eq.~(\ref{eq:pi_vec}), where it is defined as the total kinetic momentum divided by $\sqrt{N}$:
\begin{align}
  \pi^x&=\sqrt{\frac{eB}{N}}\sum_{n,n',k_x} c_{nk_x}^\dagger c_{n'k_x}\langle n,k_x|i\frac{a_{k_x}^\dagger-a_{k_x}}{\sqrt{2}}|n',k_x\rangle\nonumber\\
    &=\sqrt{\frac{eB}{N}}\sum_{n,k_x}\sqrt{\frac{n+1}{2}}(i c_{n+1k_x}^\dagger  c_{nk_x}+\text{h.c.}),\\
   \pi^y&=\sqrt{\frac{eB}{N}}\sum_{n,n',k_x} c_{nk_x}^\dagger c_{n'k_x}\langle n,k_x|\frac{a_{k_x}^\dagger+a_{k_x}}{\sqrt{2}}|n',k_x\rangle\nonumber\\
    &=\sqrt{\frac{eB}{N}}\sum_{n,k_x}\sqrt{\frac{n+1}{2}}( c_{n+1k_x}^\dagger  c_{nk_x}+\text{h.c.}).
\end{align}
Here, the ladder operator of the Landau levels are defined as 
\begin{align}
     a_{k_x}&=\frac{p_y+i(k_x-eBy)}{\sqrt{2e  B}}.
\end{align}
We notice that $\pi^x$ and $\pi^y$ satisfy the following relation:
\begin{align}
      [\pi^y,\pi^x]&=ieB \frac{N_e}{N}=i\frac{eB}{N}\sum_{n,k_x}c^\dagger_{nk_x} c_{nk_x},
\end{align}
where $N_e$ is the electron number operator. For a system with a conserved electron number $N$, this relation reduces to Eq.~(\ref{eq:pi_commu}). If we consider a single uniform cavity mode with polarization $\hat{e}_c$, the light-matter interaction can be expressed as
\begin{align}
    \frac{H'}{\omega_B}&=\sum_{n,k_x}\sqrt{\frac{\alpha\nu(n+1)}{N \omega_c L_\text{eff}}}c_{n+1k_x}^\dagger  c_{nk_x}\hat{e}_{-}\cdot\big(\hat{e}_c a+\vec{e}^{\,\ast}_c a^\dagger\big)+\text{h.c.}.\label{eq:Hint}
\end{align}
Now, we focus on solving the equation $ H'+[ H_0, S]=0$, where
\begin{align}
    H_0&=\sum_{n,k_x}\omega_B\left(n+\frac{1}{2}\right) c_{n k_x}^\dagger  c_{n k_x}+\widetilde{\omega}_c\left( a^\dagger a+\frac{1}{2}\right)\nonumber\\&+\frac{\hat{e}_c^{\,2} \alpha\nu \omega_B}{2L_\text{eff}\omega_c}(a^2+a^{\dagger 2}).
\end{align}
We first assume that $ S$ has the following form~\cite{landi2024}:
\begin{align}
   S&=\sum_{n,k_x} c_{n+1k_x}^\dagger c_{nk_x}(C_n a+D_n a^\dagger)-\text{h.c.},\label{eq:Sform}
\end{align}
where $C_n$ and $D_n$ are coefficients that need to be determined. Applying the Lemma
\begin{align}
    [ c_{n+1k_x}^\dagger c_{nk_x}, c_{n'k'_x}^\dagger  c_{n'k'_x}]= c_{n+1k_x}^\dagger c_{nk_x}(\delta_{n'n}-\delta_{n'n+1})\delta_{k_xk'_x},
\end{align}
we have
\begin{align}
    [ S, H_0]&=\omega_B\sum_{n,k_x} c_{n+1k_x}^\dagger c_{nk_x}\bigg[(-C_n a-D_n a^\dagger)\nonumber\\&+[C_n a+D_n a^\dagger, \gamma a^\dagger a+\frac{\hat{e}_c^{\,2} \alpha\nu }{2L_\text{eff}\omega_c}(a^2+a^{\dagger2})]\bigg]+\text{h.c.}\nonumber\\
    &=\omega_B\sum_{n,k_x} c_{n+1k_x}^\dagger c_{nk_x}\bigg\{\bigg[C_n(\gamma -1)-D_n\frac{\hat{e}_c^{\,2} \alpha\nu }{L_\text{eff}\omega_c}\bigg]a\nonumber\\&\bigg[C_n\frac{\hat{e}_c^{\,2} \alpha\nu }{L_\text{eff}\omega_c}-D_n(\gamma+1) \bigg]a^\dagger\bigg\}+\text{h.c.}\,.\label{eq:sh0}
\end{align}
Here, we use the notation $ \gamma=\widetilde{\omega}_c/\omega_B$ as Eq.~(\ref{eq:gammadef}). By using the relation $  [ S, H_0]= H'$ and the expression of $H'$ in Eq.~(\ref{eq:Hint}), we obtain the equations for the coefficients $C_n$ and $D_n$:
\begin{align}
    \left(\begin{array}{c}
         C_n  \\
         D_n 
    \end{array}\right)&= \sqrt{\frac{\alpha \nu(n+1)}{N\omega_c L_\text{eff}}}\left(\begin{array}{cc}
        \gamma-1 &  -\frac{\alpha\nu \hat{e}_c^{\,2}}{L_\text{eff}\omega_c}\\
        \frac{\alpha\nu \hat{e}_c^{\,2}}{L_\text{eff}\omega_c}& -(\gamma+1)
    \end{array}\right)^{-1}\left(\begin{array}{c}
        \hat{e}_-\cdot\hat{e}_c  \\
         \hat{e}_-\cdot\hat{e}_c^\ast 
    \end{array}\right).
\end{align}
For chiral cavities with polarization $\hat{e}_s=(\hat{e}_y-si\hat{e}_x)/\sqrt{2}$ ($s=\pm1$) (Eq.~(\ref{eq:epsilon_s})), the coefficients are
\begin{align}
    C_n&=-\sqrt{\frac{\alpha \nu(n+1)}{N L_\text{eff} \omega_c }}\frac{\delta_{s,+1}}{1-\gamma},\\
    D_n&=- i \sqrt{\frac{\alpha \nu(n+1)}{N L_\text{eff}\omega_c }}\frac{\delta_{s,-1}}{1+\gamma},
\end{align}
where $\delta_{ij}$ is the Kronecker delta function. Note that at the resonant point $\gamma=1$, $ H'+[ H_0, S]=0$ has no solutions if $\hat{e}_c^{\,2}=0$ and the transformation cannot be applied to the Hamiltonian. 

\subsection{Effective Hamiltonian and dressed excitations in a linearly polarized cavity}

In this section, we present the effective Hamiltonian of the quantum Hall state in a linear cavity under the SW transformation, for comparison with the results in a chiral cavity.

For a linear cavity with polarization $\hat{e}_c=\hat{e}_y$, we apply the $S$ operator with $\hat{e}_c^{\,2}=1$ to derive the effective Hamiltonian, which takes the form $ H^S\approx H_0+\frac{1}{2}[ H', S]$. The result expanded to the first order of $ \alpha \nu /L_\text{eff}$ is given by the following expressions: 
\begin{align}
    H^S&\approx H_1^S+H_2^S,\\
    H^S_1&=\omega_c(a^\dagger a+\frac{1}{2})-\frac{\alpha \nu\omega_c(a+a^\dagger)^2}{2L_\text{eff}\omega_B},\\
    H^S_2&=\sum_{n,k_x}\omega_B\left(n+\frac{1}{2}\right) c_{n k_x}^\dagger  c_{n k_x}+\frac{ \alpha \nu}{ L_\text{eff}\omega_B}\frac{ \pi^{y2}}{m_e},
\end{align}
Using the same method outlined in Appendix~\ref{app:backreaction}, we diagonalize $H^S_1$ and find that the normal frequency of the dressed photons is $\sqrt{1-2\alpha\nu/(L_\text{eff}\omega_B)}\omega_c\approx\omega_c-\alpha\nu\omega_c/(L_\text{eff}\omega_B)$. By modifying the mass in $H_2^S$ as $m_e[1+2\alpha \nu /(L_\text{eff}\omega_B)]^{-1/2}$, we obtain the Kohn mode frequency to the lowest order as $\sqrt{1+2\alpha \nu /(L_\text{eff}\omega_B)}\omega_B\approx\omega_B+\alpha \nu /L_\text{eff}$.

The above results are consistent with that found in Refs~\cite{linear2024,Rokaj2023}. 

\section{Exact results in a left-handed cavity}\label{app:exact_check}
For the cavity mode with polarization $\hat{e}_s$ defined in Eq.~(\ref{eq:epsilon_s}), the center-of-mass Hamiltonian given in the main text can be mapped to a general model of two coupled quantum harmonic oscillators,
\begin{align}
       H_\text{CM}(k_x)&=\frac{1}{2}\sum_{ij} T_{ij}P_i P_j+V_{ij}X_i X_j,\label{eq:Hcm_X}
\end{align}
by introducing the coordinates 
\begin{align}
    (X_i)&=\bigg(\frac{\pi^y_{k_x} }{\sqrt{eB }},  \frac{a+a^\dagger}{\sqrt{2}}\bigg),
\end{align}
and the dual coordinates
\begin{align}
    (P_{i})&=\bigg(\frac{\pi^x_{k_x} }{\sqrt{eB }}, \frac{i(a^\dagger-a)}{\sqrt{2}}\bigg).
\end{align}
The definitions of $\vec{\pi}_{k_x}$ can be found in the paragraph following Eq.~(\ref{eq:pi_R}). The coordinates and dual coordinates satisfy the commutation relation
\begin{align}
    [X_i,P_j]=i\delta_{ij}.
\end{align}
For an arbitrary cavity polarization $\hat{e}_c$, the mapping Eq.~(\ref{eq:Hcm_X}) can fail since there can be a mixed term between $X_i$ and $P_j$.

By applying a coordinate transformation $X_i=U_{ji}^{-1}\widetilde{X}_j$ and $  P_i=U_{ij}\widetilde{P}_j$, one can decouple the two harmonic oscillators into two normal modes $ j=\pm$, and diagonalize the Hamiltonian as
\begin{align}
    H_{\text{CM}}(k_x)&=\frac{1}{2}\sum_{ j=\pm} \omega_ j(\widetilde{P}_ j^2+\widetilde{X}^{2}_ j).
\end{align}
In terms of the new coordinates $\widetilde{X}_j$ ($j=\pm$), we introduce the ladder operators
\begin{align}
d_i&=\frac{1}{\sqrt{2}}(\widetilde{X}_i+i\widetilde{P}_i).
\end{align}
and rewrite the Hamiltonian as
\begin{align}
     H_{\text{CM}}(k_x)&=\sum_{i=\pm} \omega_i(d^\dagger_i d_i+\frac{1}{2}).
\end{align}
Thus, the eigenstates of $H_{\text{CM}}(k_x)$ with eigenenergies
\begin{align}
    E_{m_1m_2}=\sum_i\omega_i(m_i+1/2)
\end{align}
can be solved as
\begin{align}
    |m_1,m_2\rangle&=\frac{d_1^{\dagger m_1}d_2^{\dagger m_2}}{\sqrt{m_1!m_2!}}|0,0\rangle, 
\end{align}
where the ground state satisfies the equations $d_\pm|0,0\rangle=0$. 

For a chiral cavity with the polarization $  \hat{e}_s$ (defined in Eq.~(\ref{eq:epsilon_s})), the matrices in Eq.~(\ref{eq:Hcm_X}) are 
\begin{align}
    (T_{ij})&=\left(\begin{array}{cc}
         \omega_B & s  \omega_B\sqrt{\frac{\alpha\nu}{ L_\text{eff}\omega_c}} \\
     s \omega_B\sqrt{\frac{\alpha\nu}{ L_\text{eff}\omega_c}}    & \widetilde{\omega}_c
     \end{array}\right),\\
   ( V_{ij})&=\left(\begin{array}{cc}
         \omega_B & \omega_B\sqrt{\frac{\alpha\nu}{ L_\text{eff}\omega_c}} \\
     \omega_B\sqrt{\frac{\alpha\nu}{ L_\text{eff}\omega_c}}    & \widetilde{\omega}_c
     \end{array}\right),
\end{align}
where $\widetilde{\omega}_c$ is defined in Eq.~(\ref{eq:tilde_omega}). For the left-handed cavity ($s=1$) with polarization $\hat{e}_+$, the transformation to the coordinates is the following matrix that diagonalized $(T_{ij})$ because $T_{ij}=V_{ij}$:
\begin{align}
   (U_{ij}) =\left(\begin{array}{cc}
         \cos{(\theta/2)} &\sin{(\theta/2)} \\
   \sin{(\theta/2)}   &-\cos{(\theta/2)}
     \end{array}\right),\label{eq:diag_U}
\end{align}
where the angle $\theta$ is defined as
\begin{align}
    \tan{\theta}=\frac{2\omega_B}{\omega_B-\widetilde{\omega}_c}\sqrt{\frac{\alpha \nu}{\omega_c L_\text{eff}}},
\end{align}
Through the diagonalization, we have the frequencies of normal modes:
\begin{align}
    \omega_\pm&=\frac{\omega_B+\widetilde{\omega}_c}{2}\pm \frac{1}{2}\sqrt{(\omega_B-\widetilde{\omega}_c)^2+\frac{\alpha \nu}{L_\text{eff}}\frac{\omega_B}{ \omega_c}},
\end{align}
For large $L_\text{eff}$ and $\omega_B$, we find that to the first order of $\alpha \nu/L_\text{eff}$, the results by the SW transformation in Eqs.~(\ref{eq:frequency1}) and (\ref{eq:frequency2}) are consistent to the exact results.

Using Eqs.~(\ref{eq:Kubo_aa}) and (\ref{eq:Kubo_xy}), we obtain the AC conductivities for the left-handed cavity as follows
\begin{align}
    \sigma_L(\omega)&=\frac{e^2\nu}{2\pi}i\omega\bigg(\frac{\omega_+\cos{^2\frac{\theta}{2}}}{\omega_+^2-\omega^2}+\frac{\omega_-\sin{^2\frac{\theta}{2}}}{\omega_-^2-\omega^2}\bigg),\\
     \sigma_H(\omega)&=\frac{e^2\nu}{2\pi}\left[1+\omega^2\bigg(\frac{\cos{^2\frac{\theta}{2}}}{\omega_+^2-\omega^2}+\frac{\sin{^2\frac{\theta}{2}}}{\omega_-^2-\omega^2}\bigg)\right].
\end{align}
As $\omega\to0$, we have $ \sigma_H(\omega)\to e^2\nu/h$, $ \sigma_L(\omega)\to 0$.  For finite frequencies, we assume that  $\omega_B$ is much larger than all other characteristic frequencies and expand the exact results to first order in $\alpha \nu/(L_\text{eff}\omega_B)$. This leads to the exact AC conductivities reducing to the results in Eqs.~(\ref{eq:ACL}-\ref{eq:ACH})


\section{Effective hydrodynamics}\label{app:hydro}

We discuss additional details on the hydrodynamics calculation.


\subsection{The Maxwell action for a circularly polarized mode}

Let us review the derivation of the action for the cavity mode in terms of a complex amplitude. First, recall that for a single-mode cavity, with linear polarization $\hat{e}_y$~\cite{linear2024},
\begin{align}
    \vec{A}_c(\vec{x},t)=q_1(t)\hat{e}_y\phi_{k_c}(\vec{x}),\label{eq:apA}
\end{align}
where $q_1(t)$ is a real amplitude and the mode function $\phi_{k_c}(\vec{x})$ satisfies 
\begin{align}
    &\Delta\phi_{k_c} = -k_c^2\phi_{k_c}, &\int_V d^3x |\phi_{k_c}(\vec{x})|^2=1.\label{eq:psi}
\end{align}
In Ref.~\cite{linear2024} we showed that, starting from the Maxwell action, the generating functional of the electromagnetic field can be expressed as
\begin{align}
   \int \mathcal{D}q_1\exp\bigg[i\widetilde{S}[q_1]\bigg]= \int \mathcal{D}q_1\exp\bigg[i\int dt\,\frac{\varepsilon}{2}(\dot{q}^2_1-\omega_c^2 q_1^2)\bigg].
\end{align}
Then note that, introducing a complex amplitude 
\begin{align}
    q(t)=q_1(t)+iq_2(t),
\end{align}
one can express the generating functional for the cavity mode as
\begin{align}
    &\int \mathcal{D}q_1\mathcal{D}q_2\exp\bigg(iS[q]\bigg)=\nonumber\\&\int \mathcal{D}(q,q^\ast)\exp\bigg(i\int dt\,\frac{\varepsilon\omega_c}{2}q^\ast(i\partial_t-\omega_c )q\bigg),
\end{align}
as one can check by directly integrating out the imaginary part of the amplitude $q_2$, which in the linear case does not appear in the expression for the field operator \eqref{eq:apA}. To describe the circularly polarized cavity modes, we consider a linear transformation from the superposition of linearly polarized modes along $\hat{e}_{x,y}$ to the circular polarizations $\hat{e}_s$, where
\begin{align}
    \vec{A}_c&=\text{Re}[\sum_{i=x,y}\hat{e}_i (q_i(t)+q_i^\ast(t))\phi_{k_c}(\vec{x})]\nonumber\\
    &=\text{Re}[\sum_{s=\pm}(\hat{e}_s q_s(t)+\hat{e}_s ^{\,\ast} q_s^\ast(t))\phi_{k_c}(\vec{x})],
\end{align}
where
\begin{align}
    q_s=\frac{q_y+s i q_x}{\sqrt{2}}.
\end{align}
Thus, the corresponding generating functional for a single mode with polarization $\hat{e}_s$ becomes
\begin{align}
    &\int \mathcal{D}(q_s,q^\ast_s)\exp\bigg(iS[q_s]\bigg)=\nonumber\\& \int \mathcal{D}(q_s,q^\ast_s)\exp\bigg(i\int dt\,\frac{\varepsilon\omega_c}{2}q_s^\ast(i\partial_t-\omega_c )q_s\bigg).
\end{align}

\subsection{Comparison with linear polarization}
\begin{table*}[t]
\centering
\begin{tabular}{|c|c|c|}
\hline
\textbf{Property} & \textbf{Chiral Cavity} & \textbf{Linear Cavity} \\
\hline
Polarization unit vector & $\hat{e}_s = \frac{\hat{e}_y - is\hat{e}_x}{\sqrt{2}}$ & $\hat{e}_y$ \\
\hline
Dressed-photon frequency & $\omega_c - s \frac{\alpha \nu}{L_{\mathrm{eff}}}$ & $\omega_c - \frac{\alpha \nu}{L_{\mathrm{eff}}} \frac{\omega_c}{\omega_B}$ \\
\hline
Kohn mode frequency & $\omega_B ( 1 + s \nu \frac{\alpha}{L_{\mathrm{eff}} \omega_c} ) + \nu \frac{\alpha}{L_{\mathrm{eff}}}$ & $\omega_B + \frac{\alpha \nu}{L_{\mathrm{eff}}}$ \\
\hline
\parbox[t]{5cm}{Modification of longitudinal conductivities in topological limit} & $\frac{e^2 \nu}{2 \pi} \frac{\alpha \nu}{L_\text{eff}} \frac{i \omega}{\omega_c^2 - \omega^2}$ & For $\sigma_{xx}$, $\frac{e^2 \nu}{2 \pi} \frac{\alpha \nu}{L_\text{eff}} \frac{i \omega}{\omega_c^2 - \omega^2}$; For $\sigma_{yy}$, 0 \\
\hline
\parbox[t]{5cm}{Modification of Hall conductance in topological limit} & $s \frac{\alpha \nu}{L_{\rm eff} \omega_c} \frac{\omega^2}{\omega_c^2 - \omega^2}$ & 0 \\
\hline
\end{tabular}
\caption{Comparison between the results in chiral and linear cavities~\cite{linear2024}.}
\label{table:comparison}
\end{table*}

We consider the single-mode approximation,
\begin{equation}
    \vec{A}_c(t,\vec{x})=\Rep[q(t)\phi_{k_c}(\vec{x})\hat e_s],\label{eq:Acsinglemode}
\end{equation}
where the polarization vector
\begin{equation}
    \hat{e}_s=\frac{\hat{e}_y-i s\hat{e}_x}{\sqrt{1+s^2}}
\end{equation}
is a useful way to express both the linear ($s\to 0,\infty$) and chiral ($s=\pm 1$) polarizations.
The fluctuations of the cavity mode are controlled by the single-mode Lagrangian 
\begin{equation}
    L_c = \frac{\varepsilon\omega_c}{2}q^*(i\partial_t-\omega_c)q,\label{eq:Lccomplex}
\end{equation}
where $\omega_c = \frac{k_c}{\sqrt{\varepsilon\mu}}$ is the cavity frequency.

Integrating over the cavity volume reduces the coupling to the cavity mode in \eqref{eq:hydroL} to
\begin{equation}
    \frac{e}{2\pi}\int\epsilon^{\mu\nu\sigma}A_{c,\mu}\p_\nu b_\sigma=\frac{e}{4\pi}(q e_{s}^{\perp,i}+q^*e_{s}^{\perp,i*})\p_tb_{c,i},
\end{equation}
where we use notations $v^{\perp,i}=\epsilon^{ij}v_j$ for the rotation of a vector by $\pi/2$, and $f_c(t) = \int\phi_{k_c}|_H f(t,\vec x)d^2x$ for the weighted average of a function on the Hall bar, with weights given by the mode function of the cavity, $\phi_{k_c}|_H$. Here we have chosen the radiation gauge, $v^0=\vec\nabla\cdot\vec v=0$, for both the cavity and the hydrodynamic fields. It is also useful to separate the symmetric part $g_s^{ij}$ of
\begin{equation}
    (e_s^{\perp,i}e_s^{\perp,j*})=\frac{1}{1+s^2}\begin{pmatrix}
        1 &-is\\
        is & s^2
    \end{pmatrix}.
\end{equation}
With these definitions, one can integrate out the cavity fluctuations, to generate a contribution
\begin{equation}
    -\frac{2}{\varepsilon\omega_c}\left(\frac{e}{4\pi}\right)^2\int dt\p_t b_{c,i}\frac{(g_s^{ij}-is\epsilon^{ij})}{(1+s^2)(i\p_t-\omega_c)}\p_t b_{c,j}.
\end{equation}
to the hydrodynamic action.
For linear polarization along the $y$-direction ($s= 0$) this term becomes, in the frequency domain,
\begin{equation}
    -\frac{1}{2\varepsilon}\left(\frac{e}{2\pi}\right)^2\frac{\omega^2}{\omega^2-\omega_c^2}b^\ast_{c,x}b_{c,x},
\end{equation}
which is the term we report in \cite{linear2024}. The limit of linear polarization along the $x$-direction ($s\to\infty$) gives a similar term. In both cases, it follows from the symmetric component $g_s^{ij}$, and it involves only the components of the hydrodynamic field in the direction transverse to the polarization.

For the chiral cavity case, $s=\pm 1$, the resulting contribution is isotropic,
\begin{align}
    &-\frac{1}{\varepsilon}\left(\frac{e}{4\pi}\right)^2\int\frac{d\omega}{2\pi}\Big[\frac{\omega^2}{\omega^2-\omega_c^2}\delta^{ij}\nonumber\\
    &\hspace{.2\linewidth}-is\frac{\omega^3}{\omega_c(\omega^2-\omega_c^2)}\epsilon^{ij}\Big]b_{c,i}(-\omega)b_{c,j}(\omega).\label{eq:chiraltermAP}
\end{align}
Note that the sign of the chiral contribution, $\propto\epsilon^{ij}$, is given by the chirality of the cavity mode $s$. In Table.~\ref{table:comparison}, we list the main features the QH state in cavities with a circularly and linearly polarized mode to a compare the results in Ref.~\cite{linear2024}.

\bibliographystyle{apsrev4-1}
\bibliography{apssamp}

\end{document}